\documentclass[pra,showpacs,preprint,nobibnotes,nofootinbib,floatfix]{revtex4}

\usepackage{amsmath}
\usepackage{amssymb}
\usepackage{graphicx}
\usepackage{amsthm}
\usepackage{bbold}
\usepackage{float}

\DeclareMathOperator{\Tr}{Tr}
\DeclareMathOperator{\Ker}{Ker}
\DeclareMathOperator{\diag}{diag}
\DeclareMathOperator{\arctanh}{arctanh}
\DeclareMathOperator{\sgn}{sgn}

\newcommand{\R}{\ensuremath{\mathbb R}}
\newcommand{\cH}{\ensuremath{\mathcal H}}

\theoremstyle{plain}
\newtheorem{theorem}{{Theorem}}

\begin{document}

\title{Concurrence and Entanglement Entropy of Stochastic 1-Qubit Maps}

\date{\today}

\author{Meik Hellmund}
\affiliation{Mathematisches Institut, Universit{\"a}t Leipzig,
Johannisgasse 26, D-04103 Leipzig, Germany}
\email{Meik.Hellmund@math.uni-leipzig.de}

\author{Armin Uhlmann}
\affiliation{Institut f{\"u}r Theoretische Physik, Universit{\"a}t Leipzig,
Vor dem Hospitaltore 1, D-04103 Leipzig, Germany}
\email{Armin.Uhlmann@itp.uni-leipzig.de}

\begin{abstract}
  Explicit expressions for the concurrence of all positive
  and  trace-preserving (``stochastic'') 1-qubit  maps are
  presented. We construct the relevant convex roof patterns  
  by a new method. 
  We conclude   that two component optimal
  decompositions always exist.

  Our results can be transferred to $2 \times n$-quantum
  systems providing the concurrence for all
  rank two density operators as well as   lower and upper bounds
  for their entanglement of formation.

  We apply these results to a study of the entanglement entropy of
  1-qubit stochastic maps which preserve axial symmetry. Using analytic and
  numeric results we analyze the
  bifurcation patterns appearing in the convex roof of optimal decompositions
  and give results for the one-shot (Holevo-Schumacher-Westmoreland) 
 capacity of those maps.
\end{abstract}

\pacs{03.67.-a,   
03.67.Mn          
}
\maketitle

\section{Introduction}

Entanglement, together with its applications, is one of the main
features of quantum information theory \cite{NC00,petz08}. It is a
resource for new communication and computation algorithms.

A pure state $\pi = | \psi \rangle \langle \psi |$ of a quantum
system establishes quantum correlations between its subsystems,
entangling them with each other. As a general rule, the more mixed
(in the sense of majorization) the reduced density matrix
$\pi^{A} = \Tr_B\pi$ is, the stronger will be its entanglement with
the other parts. In bipartite quantum system the entanglement
is the same for either part, and we may speak of the entanglement
between both subsystems.  In addition, if one part is 2-dimensional,
the  orbits of the reduced density operators under local unitary 
transformations depend on one parameter only.

The problem of characterising entanglement becomes  more difficult
when the total system is in a general (i.~e.,~mixed) state.
There are now  quantum as well as classical correlations. Their distinction
 depends on the task in question and is, hence, not unique.
Therefore, generally, one has to choose between several entanglement
measures \cite{horodecki-2007,bengtsson06}.
Among them, the certainly most important one
is the {\it entanglement of formation} $ E_\Phi( \rho )$, discovered
by Bennett et al. \cite{BenFucSmo96}, expressing the asymptotic
number of ebits (maximally entangled qubit pairs) needed to
prepare a given bipartite state $\rho$
by local operations and classical communication (LOCC). Let
$\Phi$ denote a trace preserving positive map from one
quantum system into itself or into another one, and denote
by $S_{\Phi}(\rho)$ the von Neumann entropy of the output
$\Phi(\rho)$, given the input state $\rho$. Then we have
\begin{equation}
  \label{eq:1}
  E_{\Phi}( \rho ) = \min \sum p_j \; S  \left( \Phi( \pi_j )  \right)
\end{equation}
where the minimum is taken  over all possible
convex ($ \sum p_j = 1, p_j > 0$) decompositions of the state $\rho$
into pure states
\begin{equation} \label{eq:2}
\rho =
\sum p_j \, \pi_j, \quad \pi_j \, \hbox{ pure, i.e., }
\pi_j = | \psi_j \rangle \langle \psi_j |
\end{equation}
Let us call this quantity {\it entanglement entropy of $\Phi$}
or {\it $\Phi$-entanglement} for short.
This provides  the entanglement of formation, if
$\Phi$ is specified in Eq.~(\ref{eq:1}) to be one of the partial
traces, $\Tr_A$ or $\Tr_B$, of a bipartite quantum system. In other
words, the entanglement of formation is the $\Phi$-entanglement
with $\Phi = \Tr_B$ or $\Phi = \Tr_A$. The construction above
preserves the symmetry between both parts of a bi-partite quantum
system observed in the pure state case.

A further example for the appearance of the global
optimization problem Eq.~(\ref{eq:1}) is the HSW theorem 
of Holevo, Schumacher, and Westmoreland \cite{NC00, schumacher97,holevo98}.
It gives the one-shot or product state classical capacity
$\chi(\Phi)$ of a channel $\Phi$ by first subtracting
$E_{\Phi}(\rho)$ from $S_{\Phi}(\rho)$ and then maximizing 
this Holevo quantity $\chi^*(\rho)$ over
all input density operators:
\begin{eqnarray}
  \label{eq:3}
\chi_\Phi^*(\rho) &=& S( \Phi( \rho )) - E_{\Phi}(\rho) \\ \nonumber
\chi_{\Phi} &=&  \max_{\rho} \;\chi_\Phi^*(\rho)
\end{eqnarray}

Closed formulas for the entanglement of formation, i.e.,
analytic solutions to the global optimization problem Eq.~(\ref{eq:1})
are only known for certain classes of highly symmetric states
\cite{TV00,vollbrecht-2000}, for the $\Phi$-entanglement of a
3-dimensional diagonal channel \cite{BNU96} and for the 
exceptional case of a  pair of qubits. 
In this case of a $2\times 2$ system one knows a complete
analytic formula for the entanglement of formation. It has been
obtained first for rank two states \cite{BenFucSmo96,HilWoo97} and later
generalized to all 2-qubit states by Wootters \cite{Woo97}.

Wootters  expressed $E_\Phi(\rho)$ in terms of another entanglement measure
$C_\Phi(\rho)$, called {\it concurrence} in \cite{HilWoo97}.

Generally, one can 
replace the von Neumann entropy $S$ in Eq.~(\ref{eq:1}) by any other
unitary invariant, preferably concave, function, say $G$,
on state spaces. Substituting $G(\Phi(\pi))$  for
$S(\Phi(\pi))$ in Eq.~(\ref{eq:1}) one obtains another
entanglement measure attached to positive and trace
preserving maps. The concurrence is a measure of this kind: 
Let $\Phi$ map the
states of a quantum system into those of a 1-qubit
quantum system, i.e., a map of output rank 2. 
Then the $\Phi$-concurrence  $C_{\Phi}$ is
defined by using $G(\rho)= 2\sqrt{\det\rho}$.
To get the concurrence
of bipartite a $2 \times n$-system one sets $\Phi = \Tr_B$.
The concurrence appeared to be an interesting entanglement measure
in its own right \cite{Woo01}.
Many authors, e.g.  \cite{lozinski-2003,chen05,Fei07},
have obtained bounds for the
 concurrence of general bipartite systems.

We may now state the aim of the present paper as follows:
We study $C_{\Phi}$ and $E_{\Phi}$ for general 1-qubit
trace preserving positive maps $\Phi$. We also
exemplify in Section~\ref{examples}D how to transform our results
to rank two density operators of a $2 \times n$ quantum system.

In section \ref{roof} we explain important properties of roofs and
describe, for a positive and trace preserving map $\Phi$ from
any quantum system into a 1-qubit system, the relation
between $\Phi$-concurrence and $\Phi$-entanglement,
including entanglement of formation. 
In Section \ref{maps} we provide
an explicit expression for 
the concurrence of general
positive (stochastic) 1-qubit maps.  
We found this construction in \cite{hellmund08}.  
Afterwards we learned that a similar
result had already been obtained by Hildebrand \cite{hildebrand07,Hildebrand06}.
In this paper we elaborate on those results. The Section \ref{maps}
contain a  streamlined version of the constructions and proofs of 
Hildebrand and our unpublished work. 

Our construction of the concurrence works for all stochastic
(trace-preserving positive linear)
1-qubit maps, not only for completely positive ones. It is, therefore,
suggestive but not the topic of the present paper, to ask for
applications to the entanglement witness problem \cite{horo96}.

Section \ref{examples} is devoted to a more detailed
study of examples.
We present explicit formulas  
and intuitive pictures of the convex roof construction
for some important classes.
We start with bi-stochastic 1-qubit maps
(subsection A), followed by a short discussion of 1-qubit
channels of Kraus length two. Subsection C
explores the richness of stochastic maps commuting
with rotations about an axis. The last subsection D
explains, mainly by example, the application of our previous results 
to more general  channels
(trace preserving and completely positive maps) with 1-qubit output. 
Section \ref{entropy} is devoted to
the $\Phi$-entropy for axial symmetric stochastic maps.
We find several qualitative different phases distinguished
by the geometric pattern of their roofs. 
In Section \ref{outlook}  we shortly discuss the use of our construction 
 at  concurrence problems
for channels with higher rank.

\section{The convex roof construction}
\label{roof}
Let us elaborate on some details of the solution of the global optimization 
problem Eqs.~(\ref{eq:1},\ref{eq:2}) by the so-called convex roof
construction.
Let $G$ be a function on
the convex set $\Omega$ of density operators of a finite
quantum system. A point $\rho \in \Omega$ is a {\em roof point} of $G$
if there is an extremal convex combination Eq.~(\ref{eq:2})
such that
\begin{equation} \label{eq:1a}
G(\rho) = \sum p_j\; G(\pi_j) \; .
\end{equation}
Then the convex decomposition 
$\rho = \sum p_j \pi_j$ with $p_j > 0$ and $\sum p_j = 1$
will be referred to as {\em $G$-optimal.} Thus, if we  knew
a $G$-optimal decomposition of $\rho$, we could calculate
$G(\rho)$ from the values attained at pure states. A
roof point $\rho$ will be called {\em flat} if there
exists an optimal decomposition Eq.~(\ref{eq:1a}) 
where all values $G(\pi_j)$ are 
mutually
equal, i.e.,  $G(\rho) = G(\pi_j)$ for all $j$.

The function $G$ will be called a {\em roof} if every density operator
$\rho$ of $\Omega$ is a roof point for $G$.  Similary one defines a {\em flat
roof}  as a function $G$ for which every point $\rho$ is a flat roof point.

Let $g(\pi)$ be a function defined on the set of pure states. 
 Then $G$ is called a {\em roof extension}
of $g$ if $G(\rho)$ is a roof and $G(\pi)=g(\pi)$ for all pure $\pi$.
 On the other hand, if $G_{\rm conv}$ is a convex extension
of $g$ from the pure states to all states then
$G_{\rm conv} \leq G$ for every roof extension~$G$. The assertion
can immediately be seen from Eq.~(\ref{eq:1a}) and the very definition
of convex functions (Jensen's inequality). Since the
supremum of any set of convex functions is convex again, there
is a largest convex extension which is, however, 
not larger than any roof extension
of a function $g$.
Is this largest convex extension a roof? One knows that the
answer is ``yes'' for continuous $g$. Continuity of $g$,
together with the compactness of the set of pure states,
guaranties that the largest convex extension of $g$ is a roof
and, hence, the unique convex roof extension of $g$
\cite{BNU96,Uh98}:

\begin{theorem}
  \label{thmRoof}
Let $g(\pi)$ be a continuous real-valued function on
the set of pure states.
There exists exactly one  function $G(\rho)$ on $\Omega$ which can be
characterized uniquely by each one of the  following four properties:
\begin{enumerate}
\item $G$ is the unique convex roof extension of $g$.
\item $G ( \rho )$ is the solution of the optimization problem
  \begin{equation}
    \label{eq:55}
    G ( \rho ) = \inf_{\rho = \sum p_j \, \pi_j } \sum p_j \, g( \pi_j ).
  \end{equation}
\item $G( \rho )$ is largest convex extension of $g$ \cite{Roc70}.
\item $G$ is the smallest roof extension of $g$.
  \end{enumerate}
Furthermore, given $\rho \in \Omega$, the function $G$ is convexly linear
 on the convex hull of all
pure states $\pi$ appearing in optimal decompositions of $\rho$.

Therefore, $G$ provides a foliation of $\Omega$ into compact
leaves such that   a) each leaf is the convex hull of
some pure states and b) $G$ is convexly linear on each leaf.

If $G$ is not only linear but even constant on each leaf, it is called a 
flat roof. 
\end{theorem}
Item 1 of the theorem justifies to write ``min'' instead of ``inf''
in Eqs.~(\ref{eq:55}) and (\ref{eq:1}).

Let us apply the theorem to find out how concurrence and
$\Phi$-entanglement relate for stochastic maps $\Phi$ from an arbitrary
quantum system into a 1-qubit system. Setting 
(a la Shannon)\footnote{Our formulas are valid for arbitrary
  bases of the logarithm. The basis 2 is used for numerical 
calculations and plots of, e.g., the HSW capacity.}
$H(x_1, x_2) = - x_1 \log x_1 - x_2 \log x_2$, one has the
following:

\begin{theorem}
  \label{thmWnew}
Let $\Phi$ a stochastic map into the states of a 1-qubit system.
Denoting by $E_{\Phi}$ its $\Phi$-entanglement and by
$C_{\Phi}$ its concurrence. The function
\begin{equation} \label{Wnew0}
\xi(x) = H\left(\frac{1-y}{2}, \frac{1+y}{2}\right),
\quad 1 = x^2 + y^2
\end{equation}
is strictly convex within $-1 \leq x \leq 1$. It holds
\begin{equation} \label{Wnew}
E_{\Phi}(\rho) \geq \xi(C_{\Phi}(\rho)) \; .
\end{equation}
and this is an equality  when $\rho$ is a flat roof point of
$C_{\Phi}$.
\end{theorem}

To prove this theorem we have to collect three facts: a) For
pure states $\pi$ we have equality in Eq.~(\ref{Wnew}) and the
value of both sides is the von Neumann entropy of $\Phi(\pi)$. Hence,
both sides are extensions of $S(\Phi(\pi))$. b) The
right hand side of Eq.~(\ref{Wnew}) is  convex, see appendix A
for a proof. The left hand side is a convex roof and, hence,
not smaller than any other convex extension. This proves the
inequality Eq.~(\ref{Wnew}).
c) If $\rho$ is a flat roof point of $C_{\Phi}$, then the
same is true for any function of $C_{\Phi}$, in particular
for $\xi(C_{\Phi})$.  Therefore,
the left hand side, being a convex extension, cannot be
larger then the right one and equality holds.

In the case of the entanglement of formation of a 2-qubit system 
($\Phi=\Tr_B$)
the concurrence
is a flat roof and, hence, equality always holds in Eq.~(\ref{Wnew}).
This has been proved by Wootters \cite{Woo97}
by explicitly constructing flat optimal decompositions for
all 2-qubit density operators.

 However, already  the concurrence of a $2 \otimes 3$ 
bipartite system or of a general 1-qubit channel is not a flat roof. 
Eq.~(\ref{Wnew})  together
with the Fuchs-Graaf inequality (\cite{fuchsgraaf}, see also
\cite{hellmund08a}) for 1-qubit states 
$S(\rho) \le 2(\log 2) \sqrt{\det \rho}$   provides 
then the estimate 
\begin{equation}
  \label{eq:20}
 {\xi}(C_\Phi(\rho))\; \le\; E_\Phi(\rho)\; \le\;  \log(2)\; C_\Phi(\rho) 
\end{equation}
for all stochastic maps with 1-qubit output space, i.e., {\em for all
  stochastic maps of (output) rank 2.}

\section{Stochastic 1-qubit maps}
\label{maps}
The space ${\cal M}_2$ of hermitian 2$\times$2 matrices
$
\rho = \left(
\begin{smallmatrix}
  x_{00}     &    x_{01}  \\
 x^\ast_{01} &    x_{11}
\end{smallmatrix}\right)
$
is isomorphic to  Minkowski space $\R^{1,3}$ via
   \begin{eqnarray}
     \label{eq:52} {\bf x} = (x_0,\vec x) \quad \Longleftrightarrow\quad
 { \rho}  & = &  \frac{1}{2} (x_0 I+ \vec x \cdot \vec\sigma) \\
 & = &  \frac{1}{2}
  \begin{pmatrix}
    x_0 + x_3   & x_1 + i x_2\\
    x_1-ix_2    & x_0-x_3
  \end{pmatrix}\nonumber.
\end{eqnarray}

 We have  $ \det \rho = \frac{1}{4}(x_0^2 - x_1^2 - x_2^2 - x_3^2)
 = \frac{1}{4}{\bf x} \cdot {\bf x}$
where the dot between 4-vectors denotes the
Minkowski space inner product
 and
$    \Tr \rho  =  x_0$.   Therefore the cone of positive matrices
is just the forward light cone and
the state space $\Omega$ of a qubit, the Bloch ball,
is the intersection
of this cone with the hyperplane $V$ defined by   $x_0=1$.
In this picture 
mixed states  correspond to time-like vectors
and pure states to light-like vectors,
both normalized to $x_0=1$.

A trace-preserving positive linear map
 $\Phi: {\cal M}_2 \rightarrow {\cal M}_2$
can be parameterized as \cite{KR00}
\begin{equation}
  \label{eq:53} \Phi( \rho ) =
  \Phi \left( \frac{1}{2} (x_0 I + \vec x \cdot\vec \sigma ) \right) =
\frac{1}{2}\left(
x_0I + ( x_0 \vec t+{\bf \Lambda}\vec x) \cdot \vec \sigma\right)
\end{equation}
where $\bf \Lambda$ is a 3$\times$3 matrix and
$\vec t$ a 3-vector.

We consider the  quadratic form $q$  on ${\cal M}_2$ defined by
\begin{widetext}
\begin{equation}
  \label{eq:54}
  q^\Phi_{w}( {\bf x} ) =
4 ( \det\Phi(\rho)-w \det \rho) = \Phi({\bf x})\cdot \Phi({\bf x})
  -w\, {\bf x}\cdot {\bf x} = \sum_{i,j=0}^4 q_{ij} x_i x_j
\end{equation}
  \end{widetext}
where $w$ is some real parameter.
For pure states, i.e., on the boundary of the Bloch ball where ${\bf x}\cdot
{\bf x}=0$, the form $q({\bf x})$ equals the square of the concurrence
$C=2\sqrt{\det\Phi(\rho)}$.

Furthermore, we denote by $Q$  the linear map
 $Q: x_i\mapsto \sum q_{ij} x_j$ corresponding
to the quadratic form $q$ via polarization:
\begin{equation}
  \label{eq:515}
  Q^\Phi_w = Q^\Phi_0 - w\, \eta_{ij} =
  \begin{pmatrix}
    1-|\vec t|^2-w & -\vec t {\bf\Lambda} \\
     -(\vec t {\bf\Lambda})^T  &
w\, {\bf I}-{\bf\Lambda}^T{\bf\Lambda}
  \end{pmatrix}
\end{equation}
where $\eta_{ij}=\diag(+1,-1,-1,-1)$.
The following two statements are the central result of this section:
\begin{theorem}\label{thm1}
Let the  quadratic form $q$ and therefore the
matrix $Q$ be positive  semi-definite and  degenerate, i.e.,
$Q\ge 0$ and $\dim\Ker Q>0$. If $\Ker Q$ contains a non-zero
vector ${\bf n}$ which is
space-like or light-like, ${\bf n}\cdot{\bf n}\le0$, then
$q^{1/2}$ is a {\it convex roof}.
Furthermore, this  roof is {\it flat} if such an ${\bf n}$ exists with
$n_0=0$.
\end{theorem}
\begin{theorem}\label{thm2}
For every positive trace-preserving map $\Phi$
there  exists a unique value $w_0$ for the parameter $w$
such that the conditions  of
Theorem~\ref{thm1} are fulfilled.  Therefore, the concurrence of an arbitrary
stochastic 1-qubit map $\Phi$ is given  by
 $C_\Phi( \rho ) = \sqrt{ q^\Phi_{w_0} ( \rho ) } $.
\end{theorem}
Let us sketch the proof of Theorems~\ref{thm1} and \ref{thm2}.
The square root $\sqrt{q}$ of a positive semi-definite form $q$
on a linear space
provides a semi-norm on this space  and hence it is
convex. According to Theorem~\ref{thmRoof} we need to show
that it is also a roof, i.e., there is a foliation of the space
into leaves such that $q^{1/2}$ is linear on each leaf.
Let ${\bf n} = (n_0,\vec n)$ be a non-zero vector in $\Ker Q$.
Then for all vectors $\bf m$ we have
\begin{equation}
  \label{eq:57}
q({\bf m} +{\bf n}) = ({\bf m} +{\bf n})Q ({\bf m} +{\bf n})
= {\bf m} Q {\bf m} = q({\bf m}).
\end{equation}
  Let us start with the case where ${\bf n}$ can be chosen to have
  $n_0=0$.  Then $\vec n$ gives a direction in $V$ along which $q$
is constant. Therefore, $\sqrt{q}$ is a  {\it flat convex roof}.

\begin{figure}[h!t]
\label{fig:2}
\includegraphics[]{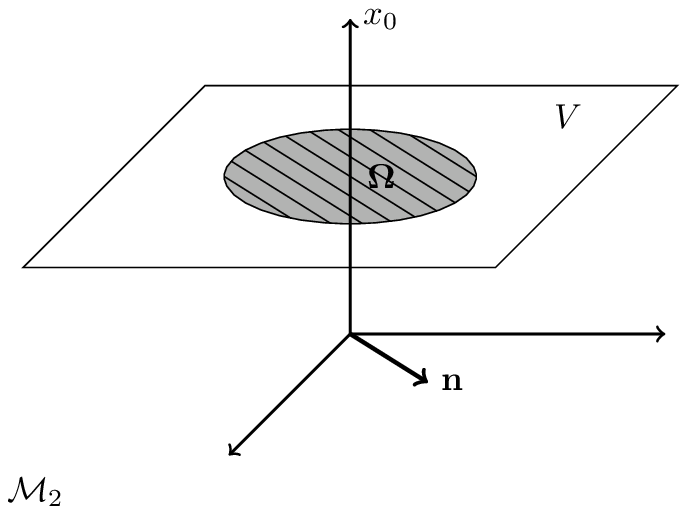}
\caption{The embedding of the Bloch ball into ${\cal M}_2$ and its
 foliation  by a flat convex roof.}
\end{figure}

Let us now consider the case where $\Ker Q$ does not contain a vector ${\bf
  n} $ with $n_0=0$. Then we have  $\dim\Ker Q=1$ and
this line intersects $V$ in
one point which we call ${\bf n}$. Every other point $\bf m$ in $V$
can be connected to the point $\bf n$ by a line lying in $V$.
Then
$q^{1/2}$ is linear along
the half-line $\R^+\ni s\mapsto s{\bf m} + (1-s) {\bf n}$ since
\begin{eqnarray}
  \label{eq:58}
  q\left(s{\bf m} + (1-s) {\bf n}\right) &=& (s{\bf m} + (1-s) {\bf n})Q
(s{\bf m} + (1-s) {\bf n}) \nonumber \\ & =&               s^2 q({\bf m})
\end{eqnarray}
This concludes the proof of Theorem~\ref{thm1}.
\begin{figure}[h!t]
\label{fig:3}
\includegraphics[]{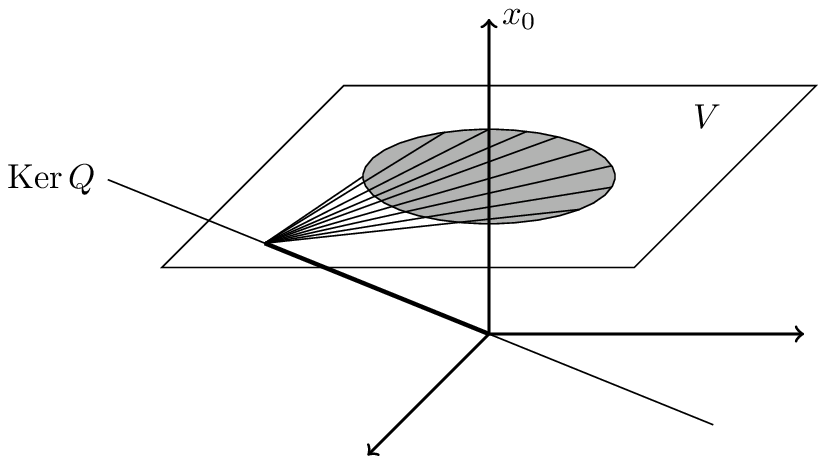}
\caption{The foliation of the Bloch ball in the case $n_0\ne 0$.}
\end{figure}
Our proof of  Theorem~\ref{thm2} presented in \cite{hellmund08} used the
Gorini-Sudarshan parametrization \cite{sudarshan76} of stochastic maps.
Here we give a shorter and more elegant 
argument following \cite{Hildebrand06,hildebrand07}.

We will consider the flow of the signature of the quadratic form 
$q=q_0-w\eta$
as function of $w\in \R$. 
It is
clear that for sufficiently large $w$ we have 
$\sgn q = \sgn(- \eta) = (+++-)$ 
whereas for large enough negative $w$ we have 
$\sgn q = \sgn(\eta) = (+---)$. 
A signature change can only occur at one of the real roots  $w_i$ of 
$\det Q= \det( Q_0-w\eta)=0$. 
The ``Minkowski metric'' $\eta$ is regular and $\eta=\eta^{-1}$.
Therefore the $w_i$ are the real eigenvalues of $\eta Q_0$ since
$\det Q= (\det \eta)\det (\eta Q_0-wI)$.  
 
Positivity of $\Phi$ implies $q^\Phi_0(\bf x)\ge 0$ for all $\bf x$ with 
$\bf x\cdot \bf x\ge 0$.  This is just the assumption of 
Yakubovich's  S-lemma from the theory of quadratic forms (see
\cite{s-lemma,Hildebrand06,hildebrand07} which ensures the existence of a
non-negative value $\hat w$ such that $q_w^\Phi$ is 
 at least 
positive semidefinite, $q^\Phi_{\hat w}\ge 0$. Then it is clear that 
all four eigenvalues of $\eta Q_0$ are real and that
 $w_1\ge \hat w\ge w_2\ge w_3\ge w_4$: There must be at least one
signature change above or at $\hat w$ and at least 3 signature changes  
below or at $\hat w$. More  signature changes are impossible 
since we have at most four real roots. There is (up to degeneracies) 
only one possible pattern of signature changes and 
 $q$ is positive and degenerate,
$\sgn q=(+,..,0)$, precisely at $w=w_1$ and $w=w_2$. It is 
positive definite for 
$w_1>w>w_2$ if $w_1\ne w_2$.  
In the case $w_1\ne w_2$ let  ${\bf n}_1, {\bf n}_2$ 
be the corresponding vectors in $\Ker Q_{w_{i}}$. Then 
 ${\bf n}_1 Q_0 {\bf n}_1 = w_1 {\bf n}_1^2$ and 
${\bf n}_2 Q_0 {\bf n}_2 = w_2 {\bf n}_2^2$. Furthermore, no nonzero vector 
can be both in $\Ker Q_{w_1}$ and $\Ker Q_{w_2}$ since $\eta$ is 
non-degenerate. So,
${\bf n}_1 Q_0 {\bf n}_1 > 
  w_2 {\bf n}_1^2$ and $ {\bf n}_2 Q_0 {\bf n}_2 > 
  w_1 {\bf n}_2^2$ (since $Q_{w_{1,2}}\ge 0$), 
providing $(w_1-w_2) {\bf n}_2^2 <0$ and 
 $(w_1-w_2) {\bf n}_1^2 >0$. Therefore, $\Ker Q$ is time-like at $w_1$ and 
space-like at $w_2$. 

In the degenerate case $w_1=w_2$, $\Ker Q$ is at least
two-dimensional. In this case, let  ${\bf n}_1, {\bf n}_2$ be two orthogonal
(in the Euclidean sense) vectors from $\Ker Q$. Then ${\bf n}_1$
and $ {\bf n}_2$ can not both be time-like  (since there is only one
time-like direction).  

This proofs the claim of Theorem~\ref{thm2}, existence of a suitable $w_0$.
It is given by $w_2$, the second largest eigenvalue of $\eta Q_0^\Phi$.

\section{Explicit examples}
\label{examples}
Let us demonstrate our construction  on some examples.
From here on  we
will sometimes denote the coordinates $x_1,x_2,x_3$ of state space
Eq.~(\ref{eq:52}) as $x,y$ and $z$. 

\subsection{Bistochastic maps or unital  channels}

Bistochastic maps preserve the center of the Bloch ball. 
We have $\vec t=0$ and the Bloch ball is pinched by $ {\bf\Lambda}
=\diag(\lambda_1,\lambda_2,\lambda_3)$.
This includes the depolarising channel $\rho\mapsto p\rho+(1-p)\frac{1}{2}I$ 
where ${\bf\Lambda}=\diag(p,p,p)$ and also the phase-damping channel 
where  ${\bf\Lambda}=\diag(p,p,1)$.
  We get 
$w=\max(\lambda_1^2,\lambda_2^2, \lambda_3^2)$  
and 
\begin{equation}
  \label{eq:9}
 C_\Phi(\rho)= q^{1/2}_\Phi(\rho) =\sqrt{ (1-w)x_0^2 +
\sum_{i=1}^3 \left( w-\lambda_i^2\right) x_i^2}
\end{equation}
which is flat in one direction since one of the terms in the sum vanishes.

Nevertheless,  this case includes channels of all Kraus lengths
between 1 and 4.

Since the roof is flat, the entanglement entropy is given by
\begin{equation}
  \label{eq:21}
E_\Phi(\rho)=
\xi\left(\sqrt{ (1-w)x_0^2 +
\sum_{i=1}^3 \left( w-\lambda_i^2\right) x_i^2}\right).
\end{equation}
The Holevo quantity
$\chi^*_\Phi(\rho)$ (see Eq.~\ref{eq:3}) is a concave function.
Since the channel is symmetric under all 3 reflections $x_i\mapsto -x_i$,
it must take its maximum, the  HSW capacity
\begin{equation}
  \label{eq:22}
  \chi_\Phi=\max_\rho \chi^*_\Phi(\rho)
\end{equation}
at the origin of the Bloch ball, $\rho=\frac{1}{2}I$. This reproduces the
well-known \cite{petz-wanabe} result
\begin{equation}
  \label{eq:23}
  \chi_\Phi= S(\frac{1}{2}I) - \xi(\sqrt{1-w})= \log(2)
  -\eta(\frac{1+\sqrt{w}}{2}) -\eta(\frac{1-\sqrt{w}}{2})
\end{equation}

\subsection{Channels of Kraus length 2 }
 A channel has Kraus length two if it can be represented as
\begin{equation}
  \label{eq:24}
  \Phi(\rho)= A^\dagger\rho A+ B^\dagger \rho B
\end{equation}
The  concurrence of
such channels has already been studied in
\cite{uhlmann05} using a quite different
approach.
 According to \cite{ruskai02}, unitary transformations can bring such a
 channel to the  form
 \begin{eqnarray}
   \label{eq:25}
   {\bf \Lambda}&=&
\diag(\cos u,\cos v,\cos u \cos v)\\
   \vec t &=&
(0,0, \sin u \sin v)
 \end{eqnarray}
which corresponds to $A=[\cos\frac{u}{2}\cos\frac{v}{2}]I +
[\sin\frac{u}{2}\sin\frac{v}{2}]\sigma_z,\; B=
[\cos\frac{u}{2}\sin\frac{v}{2}]\sigma_x -i
[\sin\frac{u}{2}\cos\frac{v}{2}]\sigma_y
$ and  we can assume $\cos u \ge \cos v$.  Then we find for the concurrence
$w=\cos^2u$ and
\begin{equation}
  \label{eq:26}
  C_\Phi^2(\rho) = y^2 (\cos^2(u) - \cos^2(v)) +
 (z\cos u  \sin v  - \cos v \sin u)^2
\end{equation}
which is positive semi-definite and independent of $x$, so we have again a
flat roof.
All channels which arise from a bipartite $2\times 2$ system with
rank-2 input states via restriction of the partial trace 
to the support space of the input state  are of length 2 and
have therefore a flat roof, in accordance with Wootters' celebrated result
\cite{HilWoo97,Woo97}.

\subsection{Axial symmetric channels}
Every positive trace-preserving linear map commuting with  rotations about
the $x_3$-axis is (modulo unitary transformations) of the form
\begin{equation}
  \label{eq:511}
  \Phi(\rho) =
\begin{pmatrix}
\alpha x_{00}  +(1-\gamma) x_{11}     & \beta x_{01} \\
\beta x_{10} &   \gamma x_{11} + (1 - \alpha) x_{00}
\end{pmatrix}.
\end{equation} with real non-negative parameters $\alpha,\beta,\gamma$.
The Bloch ball is pinched by
${\bf\Lambda}=\diag(\beta,\beta,\alpha+\gamma-1)$ and then shifted along the
$x_3$-axis by
$\vec t=(0,0,\alpha-\gamma)$.

This family includes many standard channels. Besides the 
\begin{itemize}
\item phase-damping channel (length 2, unital) for
$\alpha=\gamma=1$ and 
\item the depolarizing channel (length 4, unital) for $\alpha=\gamma,
\; \beta=2\alpha-1$
\end{itemize}
which we already considered, we also find 
\begin{itemize}
 \item the amplitude-damping channel (length 2, non-unital) for $\gamma=1,\;
\beta^2=\alpha$.
\end{itemize}

 Positivity of $\Phi$ demands
\begin{eqnarray}
  \label{eq:7}
0& \le\; \alpha,\gamma& \le\; 1, \\ \label{eq:7a}
      \beta^2 &\le\; \beta^2_{\text{max}}&=\; 1+2\alpha\gamma-\alpha-\gamma+
2\sqrt{\alpha(1-\alpha)\gamma(1-\gamma)}.
\end{eqnarray}
The first inequality guarantees that north and south pole of the Bloch ball
are not mapped to the outside, the  the second one describes the limit
 when the ellipsoid
touches the sphere at a circle.
\begin{figure}[H]
  \centering
\includegraphics[]{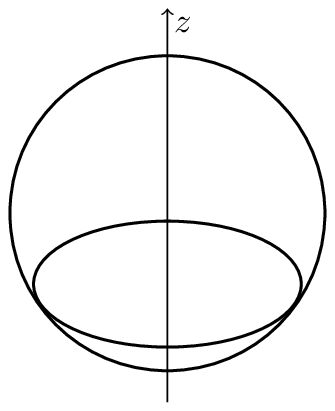}
  \caption{A map which is at the boundary of the set of positive maps.}
  \label{fig:ellip}
\end{figure}
The stronger condition of complete positivity of $\Phi$ evaluates to
\begin{equation}
  \label{eq:8}
  \beta^2 \le \alpha\gamma
\end{equation}

For the concurrence we have found the explicit expression
\begin{equation}
  \label{eq:10}
  C^2_\Phi(X) = 4(\det\Phi(X) - w \det(X))
\end{equation}
with
\begin{eqnarray}
  \label{eq:12}
  w &=& \max(\beta^2, \beta^2_c)\\
\label{eq:77}
  \text{where}\qquad \beta_c^2 &=&  1+2\alpha\gamma-\alpha-\gamma-
2\sqrt{\alpha(1-\alpha)\gamma(1-\gamma)}.
\end{eqnarray}
In the case $\beta\ge \beta_c$ we have a flat roof whose leaves are
in planes perpendicular to the $z$-axis.
\begin{figure}[H]
  \centering
\includegraphics[]{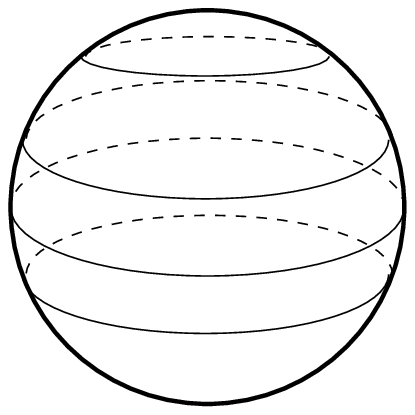}
  \caption{Leaves of the concurrence at $\beta>\beta_c$.}
  \label{fig:conc1}
\end{figure}

 In the other case we have
a one-dimensional $\Ker Q$ generated by ${\bf n}=(1,0,0,z_0)$ with
$z_0=\frac{\sqrt{\gamma(1-\gamma)}+\sqrt{\alpha(1-\alpha)}}
{\sqrt{\gamma(1-\gamma)}-\sqrt{\alpha(1-\alpha)}}$. The roof is not flat.
The leaves are straight lines meeting at the point
$z_0$
on the
$z$-axis outside the Bloch ball:
\begin{figure}[H]
  \centering
  \includegraphics[ ]{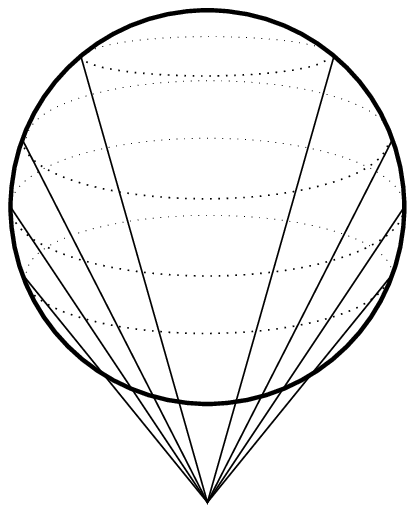}
  \caption{Leaves of the concurrence at $\beta<\beta_c$.}
  \label{fig:conc}
\end{figure}
At the bifurcation point $\beta=\beta_c$ the  concurrence is  linear
everywhere on the Bloch ball (and therefore every decomposition is optimal):
\begin{equation}
  \label{eq:13}
  C_{\beta=\beta_c} (\rho) = \left(\sqrt{\alpha(1-\alpha)}-\sqrt{
\gamma(1-\gamma)}\right) z + \sqrt{\alpha(1-\alpha)} + \sqrt{
\gamma(1-\gamma)}
\end{equation}

\begin{figure}[H]
  \centering
  \includegraphics[scale=1]{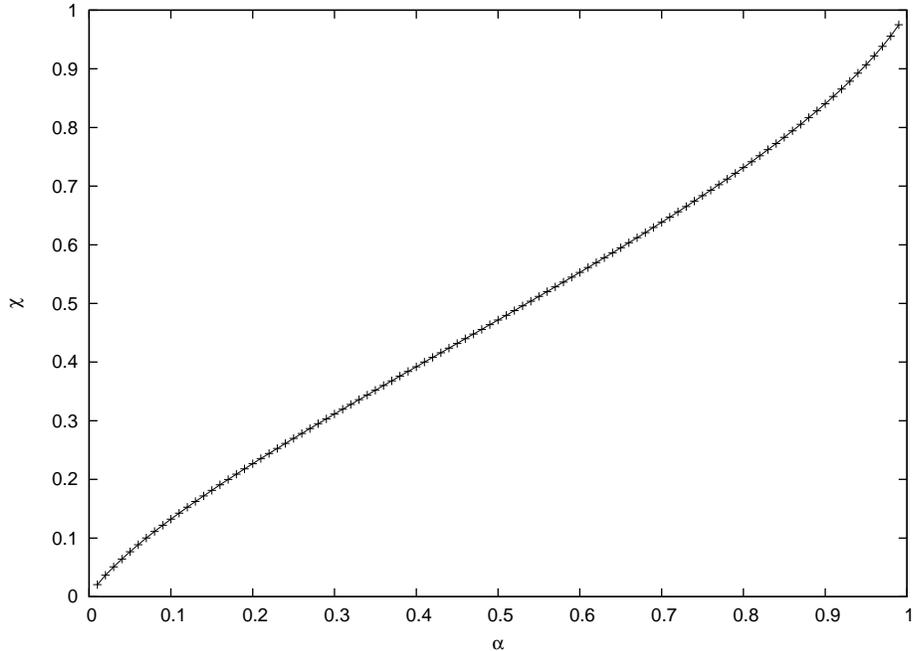}
  \caption{HSW capacity for the amplitude-damping channel as function of
   the channel parameter~$\alpha$.}
  \label{fig:ad}
\end{figure}

The special case of the amplitude-damping channel $\alpha=\beta^2,
\gamma=1$  and therefore $\beta=\beta_c=\beta_{max}$ belongs to this 
degenerate situation
with
\begin{equation}
  \label{eq:27}
  C_{AD}(\rho) = (1+z)\sqrt{\alpha(1-\alpha)} 
\end{equation}
Since this channel has length 2, this result is also a special case of
eq.~(\ref{eq:26}) for $u=-v$ with $\alpha=\cos^2 u$. 
The concave Holevo quantity must take its maximum for states on the z-axis
where we get
\begin{equation}
  \label{eq:28}
  \chi^*_{AD}(z) = \eta\left(\frac{1+z}{2}\alpha\right) + 
\eta\left(1-\frac{1+z}{2}\alpha\right) 
 -{\xi}\left((1+z)\sqrt{\alpha(1-\alpha)}\right)
\end{equation}
The equation $\frac{\partial \chi^*}
{\partial z}=0$ can be solved only numerically. The resulting capacity is
plotted in Fig.~\ref{fig:ad}. 
Similar results can be found in  \cite{dorlas}. 

Let us finally mention that we
 have no explanation for the striking similarity between eqs.~(\ref{eq:7a}) 
and (\ref{eq:77}). They differ only by the sign of the square root. So, 
$\beta^2_{max}$ derived from Fig.~\ref{fig:ellip}  
  and this $\beta^2_c$ of the roof
bifurcation   are roots of the same quadratic equation.

\subsection{The $2 \times n$ bipartite quantum system}

Here we consider $2 \times n$ systems
 $\cH = \cH^{A} \otimes \cH^{B}$, $\dim \cH^{A} = 2$.
Let $\cH_2$ be any 2-dimensional subspace of $\cH$ and $V$
a unitary mapping of $\cH^{A}$ onto $\cH_2$. Then
\begin{equation} \label{bipa1}
\rho \mapsto \Phi(\rho) = V (\Tr_B \rho) V^{\dag}
\end{equation}
is a 1-qubit channel for all density operators $\rho$ supported
by $\cH_2$. The eigenvalues of $\Phi(\rho)$ and
of $\rho^{A} = \Tr_B \rho$ are the same. Hence, by Eq.~(\ref{eq:54})
and by Theorem \ref{thm2}, we are allowed to write
\begin{equation} \label{bipa2}
C(\rho)^2 = 4 ( \det \rho^{A} - w \det \rho)
\end{equation}
for all density operators $\rho$ with support in $\cH_2$
and with a unique $w = w(\cH_2)$. Notice that this
representation does not depend on the choice of the unitary
$V$ in Eq.~(\ref{bipa1}). However, $w$ depends on the 2-dimensional subspace
$\cH_2$.

As an illustrating example we choose $n=4$ and  consider
 $\cH^{B}$ as a 2-qubit system. Then $\cH$ becomes a
3-qubit system,
$\cH = \cH^{a} \otimes \cH^{b} \otimes \cH^{c}$,
and the partial trace $\Tr_B$ from Eq.~(\ref{bipa1}) is identified
with $\Tr_{bc}$. An interesting subspace is generated
by the W and GHS  state vectors given  by
 $|W\rangle=3^{-1/3}(|001\rangle + |010\rangle + |100\rangle) $ 
and $|GHS\rangle
=2^{-1/2}(|000\rangle + |111\rangle)$. Defining
the unitary $V$ in Eq.~(\ref{bipa1}) by $V |0\rangle = |GHZ\rangle$
and $V |1\rangle = |W\rangle$, $\Phi$ can be computed to be
the 1-qubit map
\begin{equation} \label{bipa3}
\begin{pmatrix}
 x_{00} & x_{01} \\ x_{10} & x_{11}
\end{pmatrix}
 \, \mapsto \,
 \begin{pmatrix}
 \frac{2}{3} x_{00}  + \frac{1}{2} x_{11} &
\frac{1}{\sqrt{6}} x_{01} \\ \frac{1}{\sqrt{6}} x_{10} &
\frac{1}{3} x_{00}  + \frac{1}{2} x_{11}
\end{pmatrix}
\end{equation}
This an axial symmetric channel and we can read off $w = 1/6$, 
therefore,
\begin{equation}
C^2(\rho) = \frac{8}{9} x_{00}^2 +
x_{11}^2 + \frac{4}{3} x_{00} x_{11}
\end{equation}
For $\rho$  supported in our 
subspace this is equivalent to
\begin{equation} \label{bipa4}
C(\rho)^2 =
\frac{8}{9} \langle GHZ| \rho |GHZ \rangle^2 +
\langle W| \rho |W \rangle^2 + \frac{4}{3}
\langle GHZ| \rho |GHZ \rangle \, \langle W| \rho |W \rangle
\end{equation}

After this quite explicit example  we return to the more general case 
of Eq.~(\ref{bipa2}).
We rewrite the $2\times2$ determinants in  Eq.~(\ref{bipa2}) by the help 
of the characteristic equation in terms of traces:
\begin{equation}
C(\rho)^2 = 2 [ (\Tr \rho^A)^2 - \Tr ((\rho^A)^2) ]- 
2 w [ (\Tr \rho)^2 - \Tr (\rho^2) ]
\end{equation}
Polarization of this quadratic form provides (compare Eq.~(\ref{eq:54})) 
the bilinear form
\begin{equation} \label{bipa5}
q_w(\rho_1, \rho_2) = 2 (1-w)  (\Tr \rho_1) (\Tr \rho_2) 
+ 2 \left[ w (\Tr \rho_1 \rho_2) - (\Tr \rho_1^{A} \rho_2^{A}) \right]
\end{equation}
defined for all pairs  of Hermitian operators
on $\cH$. If $\rho_1$ and ÿ$\rho_2$ are supported by
the same 2-dimensional subspace $\cH_2$, and if $w$ is correctly
chosen, then $q_w$ is positive semi-definite and degenerate on
that subspace. Hence, if $C(\rho_1) = 0$, then also
$q_w(\rho_1, \rho_2) = 0$ for all $\rho_2$ supported by
$\cH_2$. In particular, if $\rho_1 = \pi_1$ is a separable
pure state and $\rho_2$ a state, we get
\begin{equation}
1-  \Tr \pi_1^{A} \rho_2^{A} =
w (1  - \Tr \pi_1 \rho_2 )
\end{equation}
It holds $\pi_1 = \pi_1^{A}\otimes  \pi_1^{B}$, as $\pi_1$ is assumed
separable.\\
If there is a second pure separable state,
say $\pi_2$, supported by $\cH_2$, one  gets
\begin{equation} \label{bipa6}
1-  \Tr \pi_1^{A} \pi_2^{A} =
w (1  - (\Tr \pi_1^{A} \pi_2^{A}) \, (\Tr \pi_1^{B} \pi_2^{B} )  )
\end{equation}
Thus, in this particular case,
the number $w$ is determined by the transition
probabilities $
\Tr \pi_1^{A,B} \pi_2^{A,B} =
|\langle\psi^{A,B}_1|\psi_2^{A,B}\rangle|^2 $
between the marginal states of $\pi_1$ and
$\pi_2$. One observes that $w$ can vary between 0 and 1 already
for subspaces generated by two separable vectors. This is a nice 
illustration of Theorem \ref{thm1}: The operator
$\pi_2 - \pi_1$ belongs to ${\rm Ker} \, Q$, and the concurrence
remains constant along the intersection of the Bloch ball
carried by $\cH_2$
with every real line of the form $\rho + t(\pi_2 - \pi_1)$.

\section{Entanglement entropy for axial symmetric
stochastic 1-qubit maps}
\label{entropy}

In this chapter we study the entanglement entropy $E_\Phi$ defined in 
Eq.~(\ref{eq:1}) for the
axially symmetric map Eq.~(\ref{eq:511}) in more detail, using
Theorem \ref{thmWnew} and numerical methods.

Our aim is an
understanding of the structure of the foliation of the Bloch ball 
provided by the convex roof construction. This foliation encodes the optimal
decompositions Eq.~(\ref{eq:2}) for all states. The foliation changes with the
channel parameters.  In most of the $(\alpha,\beta,\gamma)$ parameter space
all states have an optimal decomposition into two pure states. In a small
region of the parameter space we find optimal decompositions of length~3.
  We characterize the bifurcation structure of this ``phase transition''
and its position in parameter space. 

There exist quite a lot numerical and analytical work about the HSW
capacity of 1-qubit channels, e.g., 
\cite{cortese-2002,berry,Li-zhen}
where the optimal decomposition of the optimal state is considered. In
contrast, we consider the optimal decomposition of all states.

\subsection{Some degenerate channels}

\paragraph{$\alpha=\gamma$}
In this case the channel is unital and has therefore a flat convex roof
for the concurrence. We have $\beta_{\text{max}}=1$ and
$\beta_{c}^2 = (2\alpha-1)^2$, so
we find   $w=\max((2\alpha-1)^2,\, \beta^2)$. The concurrence
$C$, and hence  $E_\Phi$ too, are
constant either (in case of  $ (2\alpha-1)^2> \beta^2$) on concentric
cylinders around the $z$-axis $E_\Phi=E_\Phi(x^2+y^2)$ or on
planes perpendicular to the $z$-axis $E_\Phi=E_\Phi(z)$.

\paragraph{$\alpha+\gamma=1$}
In this case the range of the channel is degenerate, being 
a 2-dimensional ellipse
orthogonal to  the $z$-axis. Furthermore, $\beta_c=0$ and therefore
$w=\beta^2$. We get again a flat roof. $C_\Phi(z)$ and hence $E_\Phi$, too,
are constant on
planes perpendicular to the $z$-axis.

\subsection{The general case $(\alpha-\gamma)(\alpha+\gamma-1)\ne 0$}

We did extensive numerical studies of the global minimization problem
of the entanglement entropy Eq.~(\ref{eq:1}),
guided by and compared to analytic studies of special
cases. The following overall picture emerged: There are 3 different phases.
For fixed values of $\alpha$ and $\gamma$, we have at large values of
$\beta$ a phase (phase I) where the entanglement depends only on $z$. 
By decreasing
$\beta$, we reach phase II where a cone with apex at the north pole appears.
States in the cone have optimal decompositions of length 3.
The opening angle of the cone decreases and for small enough
$\beta$ we reach phase III, where again all optimal decompositions have
length 2.
\begin{figure}[H]
  \centering
\includegraphics[]{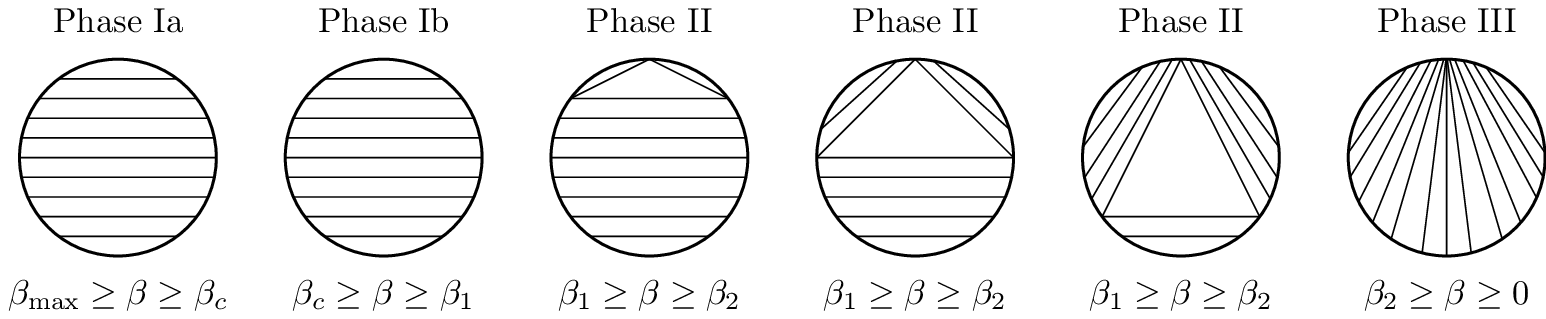}
  \caption{Leaves of the  foliation of the entanglement entropy. The $z$
    axis points upwards.}
  \label{fig:entp}
\end{figure}

Of course we have a flat entanglement roof as long as we have a flat
concurrence roof (phase Ia). But the flat phase for the entanglement
extends to even lower values
of $\beta$, where the concurrence is not longer flat (phase Ib)!

For phase III  let us remark that the leaves form cones with their apex on
the $z$-axis outside the Bloch ball. But different to the Phase II of the
concurrence (compare Fig.~\ref{fig:conc}) they do not  intersect at the
same point on the $z$-axis.

The above picture and the equations for $\beta_1$ and $\beta_2$ below are
valid in the case
\begin{equation}
  \label{eq:16}
(\alpha-\gamma)(\alpha+\gamma-1)>0.
\end{equation}
For the opposite case, turn the pictures upside down ($z\rightarrow -z$)
and exchange $\alpha \leftrightarrow \gamma$ in the equations below for
$\beta_1$ and $\beta_2$.

The bifurcation points $\beta_1$ and $ \beta_2$ between the 3 phases can be
calculated analytically. Let $s(\cos(\phi))$ denote the entropy
$S(\Phi(\pi))$ for the pure
state $\pi=(\sin(\phi),0,\cos(\phi))$. Then the bifurcation point  $\beta_1$
can be found by comparing the competing decompositions
$E_1= \frac{1}{3} s(1) + \frac{2}{3}
s(\cos(\phi))$
 with
$E_2 = s(\frac{1}{3}+\frac{2}{3}\cos(\phi) )$. We expand
$E_1(\phi)-E_2(\phi) = g(\alpha,\beta,\gamma) \phi^2  + O(\phi^3)$ and
get $\beta_1$ as the root of $g(\alpha,\beta,\gamma)=0$.

Using the abbreviations $x=2\alpha-1, y=2\gamma-1$ we find

\begin{equation}
  \label{eq:15}
  \begin{split}
  \beta_1^2 =&
\frac{x}{2(x+(x^2-1)\arctanh(x)}
 \left(x^2 + x y + ( x^2-1 ) y \arctanh(x) -  \right.\\ &\left.
     \sqrt{( 1-x^2 )  \arctanh(x) (
      x^3 - x y^2 - ( x^2-1 ) y^2 \arctanh(x)) }\right)
  \end{split}
\end{equation}

Analogously, we obtain $\beta_2$ by comparing the decompositions
$E_1= \frac{1+\cos(\phi)}{2} s(1) + \frac{1-\cos(\phi)}{2} s(-1)$ and
$E_2 = s(\cos(\phi))$ around $\phi=\pi$:

\begin{equation}
  \label{eq:14}
  \beta_2^2 =
y\,\frac{(1+x)\log(1-y) + (1-x) \log(1+y)- (1+x) \log(1+x) -(1-x)\log(1-x)}
{2(\log(1-y) -\log(1+y))}
\end{equation}

\subsection{Phase diagram}

The following figure shows the phases in the $\beta,\gamma$-plane for
$\alpha=0.8$. The upper boundary is given by the positivity condition,
Eq.~(\ref{eq:7a}). The boundary between phases Ia (entanglement and
concurrence have flat roofs) and Ib (only entanglement has flat roof) is
given by Eq.~(\ref{eq:77}). Phase II is bounded by Eqs.~(\ref{eq:15}) and
 (\ref{eq:14}).

\begin{figure}[H]
  \centering
\hspace*{-1cm}\includegraphics[scale=1.4]{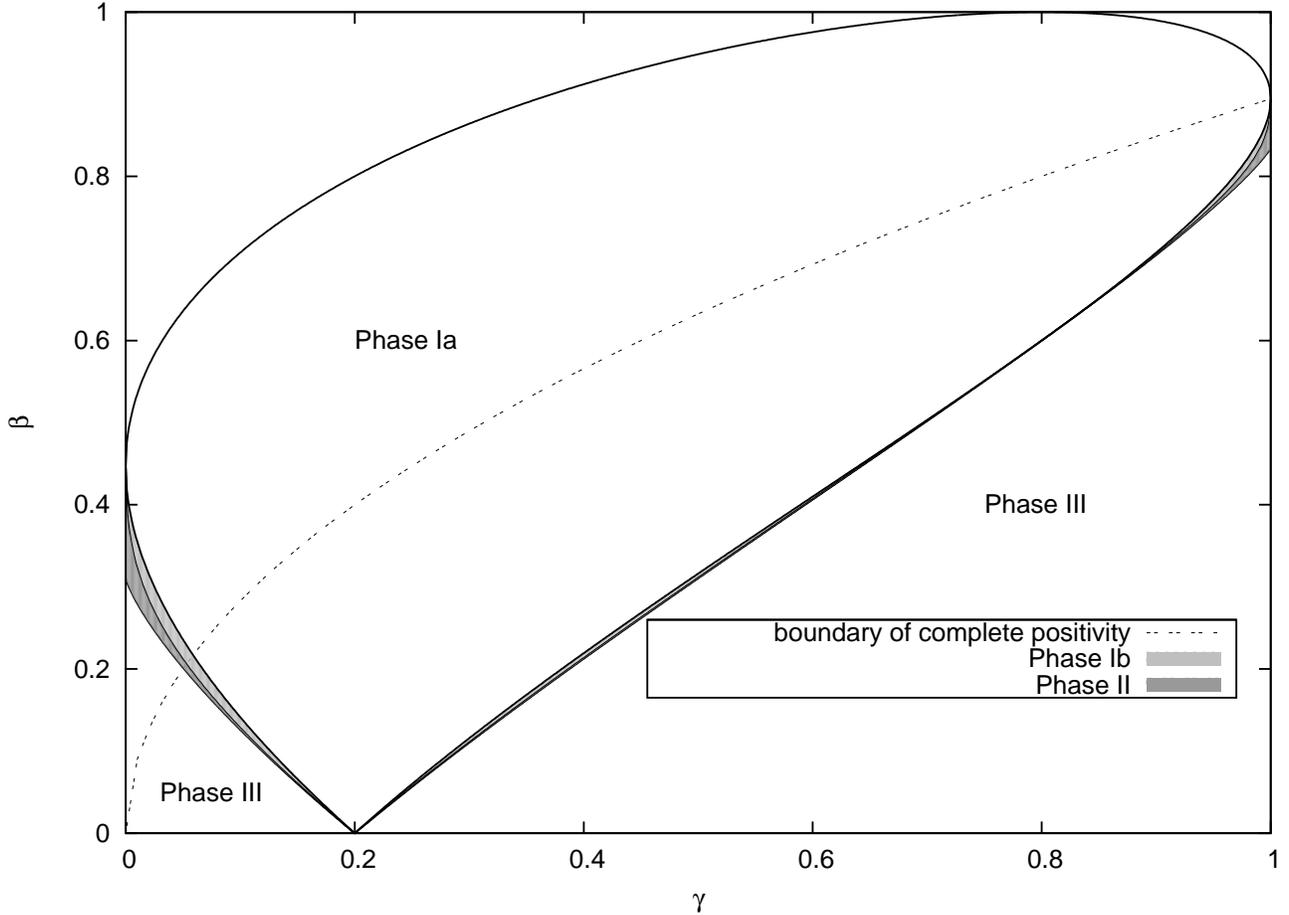}
  \caption{Phase diagram in the $(\gamma,\beta)$-plane for $\alpha=0.8$.}
  \label{fig:phase1}
\end{figure}

The phase II region where length 3 optimal decompositions exist
as well as the phase Ib are  quite
small but they exist everywhere outside  the degenerate points
where either $\alpha+\gamma=1$ or $\alpha=\gamma$.

\subsection{One-shot (HSW) capacity}
\begin{figure}[H]
  \centering
  \includegraphics[scale=1]{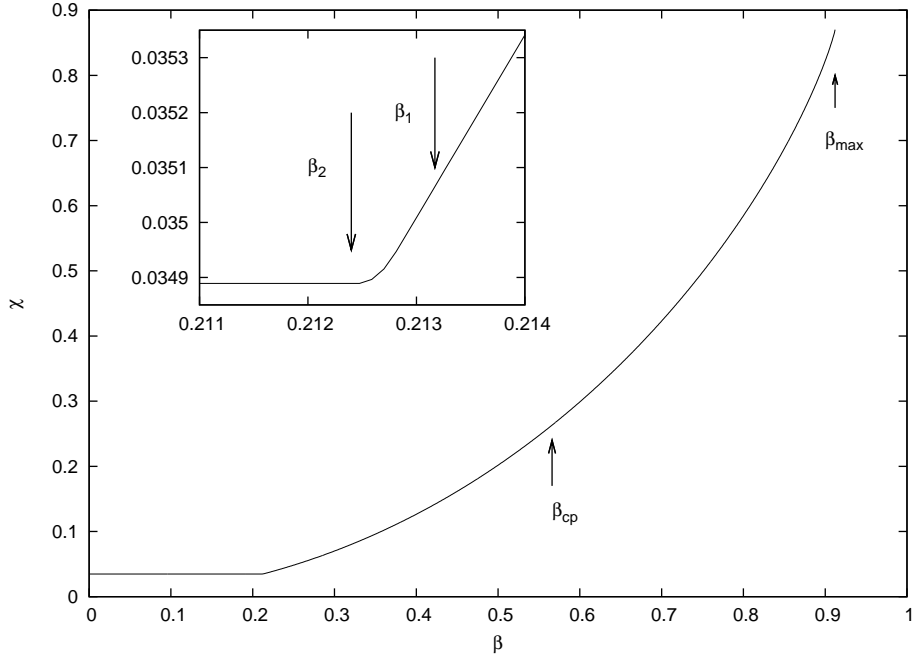}
  \caption{The HSW capacity as function of the channel parameter $\beta$
    at $\alpha=0.8,\;\gamma=0.4$. The inset shows the small region where the
  transition between phases I, II and III takes place.}
  \label{fig:holax}
\end{figure}

The Holevo quantity will take its maximum for a state on the $z$-axis.
Its numerical calculation is highly simplified by taking the foliation
structure  into account. 
We show in Figure~\ref{fig:holax} the $\beta$ dependence of this maximum,
i.e., the HSW capacity,
for  fixed values
of $\alpha$ and $\gamma$.  The maps are
positive for $\beta\le \beta_{max}$, completely positive for $\beta\le
\beta_{cp}$.  The  values  $\beta_1$ and $\beta_2$ indicated in the figure
separate 
the phases I, II and III. In phase III the capacity is  independent
of $\beta$.

\section{Concurrence for channels 
with higher input or output rank}
\label{outlook}

Our method provides a complete solution for the concurrence of 
trace-preserving positive maps  
of input and output rank~2. 
How  could one possibly overcome the input rank two (or 1-qubit map)
restriction? The following problem may be of interest:
Assume $\sum p_j \pi_j$ is an optimal decomposition
for the concurrence of a $2 \times n$ system. Every pair
$\pi_j$, $\pi_k$ of different pure states is supported by a
2-dimensional Hilbert space $\cH_{jk}$. Hence there is a
number $w_{jk} = w(\cH_{jk})$
defining  the concurrence for density operators supported
by $\cH_{jk}$ according to Eq.~(\ref{bipa2}). Which restrictions
on the set of all $w_{jk}$ arise from the optimality of the decomposition?

Another issue is the generalization to higher output ranks. 
Rungta et al \cite{RBCHMW} proposed to replace the determinant 
$\det \rho$ by the 
second elementary symmetric function of the eigenvalues,
$   C_\Phi(\pi) = 2 \sqrt{e_2(\Phi(\pi))}$. While the 
square root of $e_2$ is concave, one might find a value for $w$ making the 
expression 
\begin{equation}
  \label{eq:5}
 2\sqrt{  e_2(\Phi(\rho)) -  w\, e_2(\rho)}
\end{equation}
a convex extension of 
$2 e_2(\Phi(\pi))^{1/2}$, $\pi$ pure. In these cases, the expression
 Eq.~(\ref{eq:5}) is a lower bound for the $\Phi$-concurrence.
An example is the  diagonal map $D_m$ in any dimension $m$
which cancels the off-diagonal elements. Denoting the matrix
elements of $\rho$ by $x_{jk}$, this recipe results in
\begin{equation} \label{hdim2}
C_D(\rho) \geq 2 (\sum_{j < k} |x_{jk}|^2)^{1/2}
\end{equation}
Another example is the following family of 
indecomposable Choi maps of a $3\times 3$ system:
\begin{equation} \label{hdim3}
\rho\; \mapsto\; \Phi[\mu](\rho)\; =\;
\frac{1}{1+\mu}
\begin{pmatrix}
x_{00} +\mu  x_{22} & -x_{01} & -x_{02} \\
 -x_{10} & x_{11} + \mu x_{00} & -x_{12} \\
-x_{20} & -x_{21} & x_{22} + \mu x_{11}
\end{pmatrix}.
\end{equation}
$\Phi[\mu]$ is trace-preserving, positive and indecomposable  for
$\mu\ge1$. 
The map $\Phi[1]$ is extremal in the set of positive maps. 
Here our recipe provides the bound 
\begin{equation} \label{hdim4}
C_{\Phi}(\rho)^2 \ge 
\frac{4\mu}{(1+\mu)^2}\left[
(x_{00} + x_{11} + x_{22})^2 + (\mu-1)\left(
|x_{01}|^2 + |x_{02}|^2+ |x_{12}|^2 
\right)\right],
\end{equation}
a  positive  semi-definite quadratic form  in the
matrix entries. In the special case $\mu=1$ our recipe provides 
 an exact though 
highly degenerate answer:
$\Phi[1]$ maps all pure states of the $3\times 3$ system to mixed states 
with the same $\Phi$-concurrence and therefore the $\Phi$-concurrence is
constant everywhere, $C_\Phi(\rho)= 1.$

\section{Conclusions}
\label{conclusions}

We have explained a way to get concurrences of stochastic
1-qubit maps and of rank two states in $2 \times n$ quantum
systems. The methods is attractive by its simplicity, providing
a large area of applications. The new methods is different
from that of Wootters \cite{Woo97} and of \cite{uhlmann00}
which is based on conjugations. 

The advantage of the new
methods is its applicability to roofs which are not flat.
Only a small subset of the stochastic 1-qubit maps actually has
a $\Phi$-concurrence which is a flat roof.
For a general 1-qubit map the  concurrence is
real linear on each member of a unique bundle of straight
 lines crossing the Bloch ball. The bundle consists either
 of parallel lines or the lines meet at a pure state, or
 they meet at a point  outside the Bloch ball. Furthermore,
 $C_{\Phi}$ turns out to be the restriction of a Hilbert
 semi-norm to the state space.

For the special case of an axial symmetric 1-qubit channel we 
presented a throughout study of the $\Phi$-entanglement. 
Here the structure of the optimal decomposition of states
can be quite different depending on
the channel parameters. There is a phase where all optimal decompositions
have length 2 and are flat, a phase where states with optimal decompositions
of length 3 exist, forming a cone in the foliation of the Bloch ball,
and a phase where all optimal decompositions are of length 2 but not flat. 
We found  explicit formulas for the bifurcation points which separate the
phases.  
Interestingly, there exists a region in the space of 1-qubit maps where 
the $\Phi$-entanglement is flat despite the fact that 
 the $\Phi$-concurrence is not flat.

Our  method  of finding optimal decompositions for the concurrence 
works  perfectly for rank two density operators only. For higher rank states 
it provides lower bounds.
It is a challenge to
find an algorithm, if existing, which  combines  the merits
of this approach and the conjugation based one.

\appendix
\section{}
The function defined in Eq.~(\ref{Wnew0})
\begin{equation}
\xi(x) = H\left(\frac{1-y}{2}, \frac{1+y}{2}\right),
\quad 1 = x^2 + y^2
\end{equation}
is defined on $-1\le x \le 1$ and does not depend on the sign of
$x$. It is strictly convex since
\begin{eqnarray}
  \label{eq:18}
  \xi''(x) &=&\frac{1}{2y^3}\ln \frac{1+y}{1-y} - \frac{1}{y^2} \\
  &=& \frac{1}{y^2}\left(
    \frac{y^3}{3}+\frac{y^5}{5}+\frac{y^7}{7}+\cdots\right) \;>\; 0
\end{eqnarray}
Therefore, $\xi$ is the supremum of a family of functions
$ax + b$. Inserting a convex function $C(\rho)$ with values
$-1 \leq C \leq 1$ represents $\xi(C)$ by a supremum of
convex functions $aC + b$. This proves the convexity of
$\xi(C(\rho))$ as a function of $\rho$.


\end{document}